\documentclass[11pt]{article}%[12pt,a4paper]{article}%[\documentclass[mathserif,compress]{article}%{beamer}%[useAMS,mathserif,usenatbib,compress]{article}%[twocolumn,twoside nobibnotes,%
\usepackage{psfrag,color}
\usepackage{epsfig}
\usepackage{graphicx}
\usepackage{times}
\usepackage{url}
\usepackage{cmap}
\usepackage[TS1,T2A]{fontenc}
\usepackage[koi8-r]{inputenc}
\usepackage{natbib}
\usepackage[english]{babel}%,russian
\bibpunct{(}{)}{;}{a}{}{,}
\newcommand{\kpc}{\ensuremath{\rm kpc}}

\newcommand{\kms}{\mbox{km~s}^{-1}}

\renewcommand{\mag}[1]{^{\rm m}\!\!\!#1\,}
\newcommand{\Msun}{\ensuremath{\rm M_\odot}}
\newcommand{\Lsun}{\ensuremath{\rm L_\odot}}

\newcommand{\objsev}{\rm Cyg~OB2~\textnumero7~}

\newcommand{\apjjj}{\ensuremath{\rm Astrophys. J.}} %  Astrophysical Journal
\newcommand{\href}[1]{\url{#1}}
\newcommand{\aaa}{Astronom. and Astrophys.}
\newcommand{\apss}{Astrophys. and Space Sci.}
\newcommand{\mnras}{Monthly Notices Roy. Astronom. Soc.}
\newcommand{\pasp}{Publ. Astronom. Soc. Pacific}
\begin{document}
\title{Medium resolution spectroscopy of the supergiant
O3I$f_*$ \ \objsev}
\author{O.V.~Maryeva$^{1}$, R. Ya. Zhuchkov$^{2}$\\[2mm]
\begin{tabular}{l}
$^{1}$ Special Astrophysical Observatory, Nizhnij Arkhyz 369167, Russia
 \\ \texttt{ olga.maryeva@gmail.com}\\
$^{2}$ Astronomy and  geodesy department of Kazan (Volga region) Federal University,\\ Kremlevskaya str., 18, Kazan, 420008, Russia.\\
\end{tabular}
}
\date{}
\maketitle

\begin{abstract}
      We examine the feasibility of using medium resolution spectra for determining the
      parameters of atmospheres of hot stars by means of numerical simulations. We chose
      the star \objsev as a test object and obtained its spectrum ($\lambda/\Delta\lambda=2500$)
      with Russian-Turkish RTT150 telescope.  The {\sc CMFGEN} code was used to construct
      a model of the atmosphere of \objsev. For the first time we have detected the NIV
      $\lambda\lambda 7103.2-7129.2$lines in the spectrum of this star and used them to
      determine the physical parameters of the wind. The rate of mass loss measured using
      the H$\alpha$ line exceeds the loss rate measured using lines formed in the wind.
      This indicates that the wind is nonuniform, apparently due to rotation.

\end{abstract}


\section{Introduction}
      Since the 1960's, space based observations made it possible to detect lines with
      a P~Cyg profile in the ultraviolet spectra of O-type stars, which indicated the
      presence of supersonic winds. This discovery allowed to estimate the mass loss
      rate due to a stellar wind. Further studies showed that massive stars ($>50~\Msun$)
      lose a substantial fraction of their mass (almost half) in the form of wind while
      on Main Sequence. Later, wind lines (or details of their profiles) were also
      discovered in high-resolution spectra in the optical wavelength range obtained
      at ground-based observatories.

      The next important step for astrophysics was the discovery of a new method for
      determining distances using the so-called wind momentum-luminosity relation (WLR)
      \citep{Kudritzki}. However spectral monitoring of selected O-supergiants has shown
      that the winds from hot stars are inhomogeneous and vary with time (see \citet{Owocki} for details).
      Therefore, to determine their luminosities it is important to obtain averaged,
      statistically significant characteristics of details of the wind line in spectra
      of rare O-supergiants. Significant fraction of these stars can only be studied
      with medium-resolution spectroscopy. Moreover blue parts of their spectra may
      be attenuated as a result of interstellar and circumstellar extinction.

      In this paper we test whether is it possible to reliably determine the atmospheric
      parameters by modeling medium-resolution spectra. We have chosen the supergiant
      O3I$f_*$ \objsev as a test object. Blue part of its spectrum is  significantly
      absorbed, $A_v=5.4$ \citep{KiminkiAv}. The star belongs to the Cyg OB2
      association \citep{KlochkovaNo12,KlochkovaCygOB2}, so we can obtain an independent
      estimate of its luminosity.

      In the next section we present observational data and data reduction process. In
      Section~\ref{sec:results} we describe model calculations and discuss results,
      comparing these with previous work. The conclusions are considered in Section~\ref{sec:disc}.

%%%%%%%%%%%%%%%%%%%%%%%%%%%%%%%%%%%%%%%%%%%%%%%%%%
\section{Observations and data reduction}\label{sec:obs}

      Observations of \objsev have been performed on February-March 2012 on
      the 1.5-m Russian-Turkish RTT-150 telescope at the T\"ubitak National
      Observatory, located on Mt.~Bak${\rm \acute y}$rl${\rm \acute y}$tepe
      (Turkey). The spectrum was obtained over a wide range of wavelengths
      (4200-8000 \AA\AA) using the TFOSC (T\"ubitak NationalFaint Object
      Spectrograph and Camera\footnote{\href{www.tug.tubitak.gov.tr/rtt150_tfosc.php}})
      instrument at the Cassegrain focus. The spectral resolution is
      $\lambda/\Delta\lambda=2500$. In the overall spectrum the signal-to-noise
      ratio ${\rm S/N}=100$ in the blue part (5000 \AA) and 200 in the red part
      (7000 \AA). As noted above, the spectrum of this object is strongly absorbed
      at shorter wavelengths. Reliable modelling of faint lines requires a fairly
      high ${\rm S/N}$ ratio (>100) which, in our case, was realized only in the
      red for $ \lambda> 5000$~\AA. Below we study the spectrum in this region.
      Data were reduced  using the DECH software package .

      The CIV $\lambda\lambda 5801.3, 5812$, NIV $\lambda\lambda 6214, 6219 $ emission
      lines are present in the spectrum. For the first time the NIV $\lambda\lambda 7103.2
       - 7129.2$ lines were detected in the spectrum of \objsev,   they involve a transition
      from the $1s^22s3d$ to the $1s^22s3p$ state. These emission lines are typical for
      spectra of early Wolf-Rayet (WR) stars and are used for a spectral classification of
      nitrogen-rich WR (WN). From published spectra of O-stars in the 7000-8000 \AA\AA \
      range we may conclude that the NIV  $\lambda\lambda7103.2 - 7129.2$ lines are present
      only in the spectra of O2-O5 supergiants. Modeling shows that these lines are formed
      only when effective temperature $T_* > 38000 K $ \citep{me}.

%%%%%%%%%%%%%%%%%%%%%%%%%%%%%%%%%%%%%%%%%%%%%%%%%
\section{Modeling results}\label{sec:results}

      We have used the {\sc CMFGEN} atmosphere code %developed by J. D. Hillier
      \citep{Hillier5} to determine the physical parameters of the atmosphere of \objsev.
      This code solves radiative transfer equation for objects with spherically symmetric
      extended outflows using either the Sobolev approximation or the full comoving-frame
      solution of the radiative transfer equation. {\sc CMFGEN}  incorporate a line
      blanketing, the effect of Auger ionization and clumping. Every model is defined
      by a hydrostatic stellar radius $R_*$, luminosity $L_*$, mass-loss rate $\dot{M}$,
      filling factor $f$,  wind terminal velocity $v_\infty$, stellar mass M, and by the
      abundances $Z_i$ of included elementary species.

      Using the model of the star AV~83 (O7 $Iaf^{+}$) calculated by \citet{HillierAV83}
      as the seed model we have adjusted its parameters to reproduce the observed spectrum
      of \objsev and gradually changed the parameters of the model ($L_*$, $R_*$ and $\dot{M}$).

      In our calculations we assumed that:
\begin{itemize}
      \item[-] the volume filling factor $f_\infty$ is equal to $0.1$, as in the initial model;
      \item[-] $\beta-$law for wind  velocity and $v_\infty=3080~\kms$ (a value taken from \citet{HerreroUV});
      \item[-] H, He, C, N, O, S, Si, P, and Fe were included in calculations;
      \item[-] the abundances of S, Si, P, and Fe are solar abundances;
      \item[-] the abundances of H, He, C, N, and O are the same as in the initial model,
               ($[{X(N)}/{X(N)_\odot}]\sim3$, $[{X(C)}/{X(C)_\odot}]=0.08$, $[{X(O)}/{X(O)_\odot}]=0.09$).
\end{itemize}

      We used photometric data for an exact determination of the luminosity. The model flux was recalculated
      for the distance of Cyg OB2 association (1.5 \kpc, \ \citet{Dambis}). Then we added interstellar
      absorption using the IDL program  { FM-UNRED} (W. Landsman) which uses the absorption curves calculated
      by Fitzpatrick [12]. The value $A_v=5.4$ is taken from  \citet{KiminkiAv}. Then, the simulated spectra
      were convolved with V-band sensitivity filters. The resulting fluxes were converted to magnitudes and
      compared to the photometrical data  ($V=10.\mag~5$, Simbad data base).

      As a result, we constructed a best-fit model. Its parameters are: $L_*=(1.1\pm0.1)\times10^6\Lsun$,
      $R_* = 16.5 R_{\odot}$ , $T_* = 44\pm  1 kK$ and $\beta= 1$. $R_*$ is the radius of the star,
      corresponding to the inner boundary of the atmosphere lying approximately at  $\tau \sim 20$,
      and $T_*$ is the effective temperature at radius $R_*$, which is related to the luminosity by
      $L_*=4\pi R^2_{*} \sigma T_{*}^4$. The mass loss rate is $\dot{M}_{cl}=(3\pm0.5)\times 10^{-6}
      M_{\odot} /\mbox{year}$. The unclumped mass loss rate ($M_{uncl}$) is related to the clumped rate ($M_{cl}$) by
$
\dot{M}_{uncl}=\dot{M}_{cl}/\sqrt{f}
$ \ .

\begin{figure*}
\centering
\epsfig{file=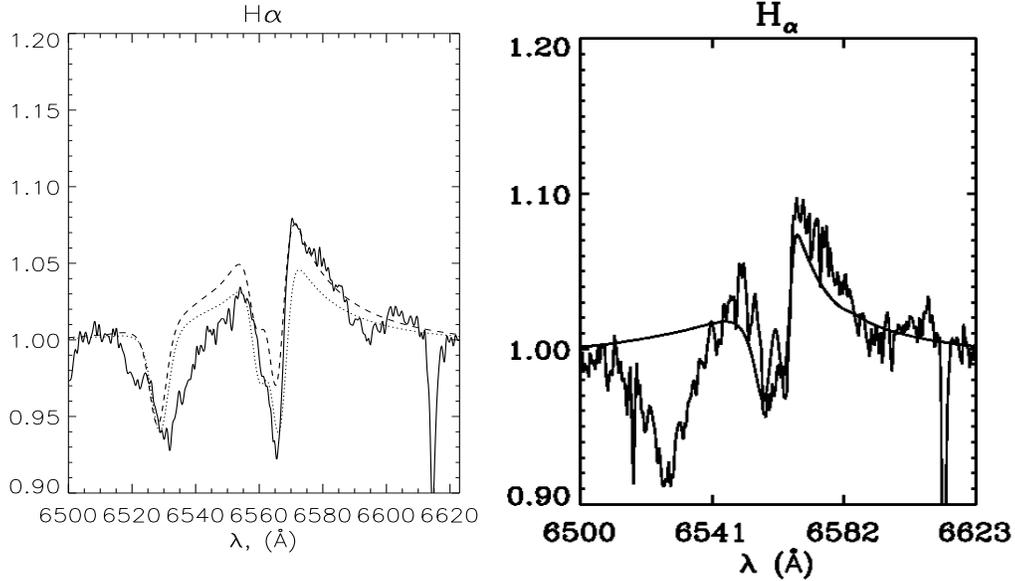,width =0.9\textwidth}%\epsfig{file=Halpha1,width =0.5\textwidth}
\caption{{\bf Left:} A comparison of the observed H$\alpha~+~$HeII$\lambda~6560$
      (solid line) spectrum with the models reported here. The dotted line is a model
      with $\beta=1$ and $\dot{M}_{cl}=2\times 10^{-6} M_{\odot}/\mbox{year}$ and the
      dashed curve, with $\beta=1$ and $\dot{M}_{cl}=2.5\times 10^{-6} M_{\odot}/\mbox{year}$.
      Another line, HeII$\lambda~6527$, is seen to the left of H$\alpha$ and DIB
      $\lambda6613$ to the right. {\bf Right:} A comparison of the theoretical and observed
      profiles of H$\alpha~+~$HeII$\lambda~6560$  from \citet{Herrero2002}.}
\label{fig:halphamodel}
\end{figure*}

      Figure~\ref{fig:halphamodel} shows comparison between the calculated and observed
      H$\alpha~+~$HeII$~\lambda 6560$ profiles. The rotational velocity of the star is
      $V \sin{i}=105~\kms$ \citep{Herrero2002}. In order to account for the star's rotation
      and the spectral resolution of the instrument ($\Delta\lambda=2$\AA), we convolved the
      calculated spectrum with a Gaussian of FWHM=2.65\AA.

%888888888888888888888888888888888 figure 1 88888888888888888888888
%8888888888888888888888888888888888 figure 1 8888888888888888888888

\begin{figure*}
\centering
\epsfig{file=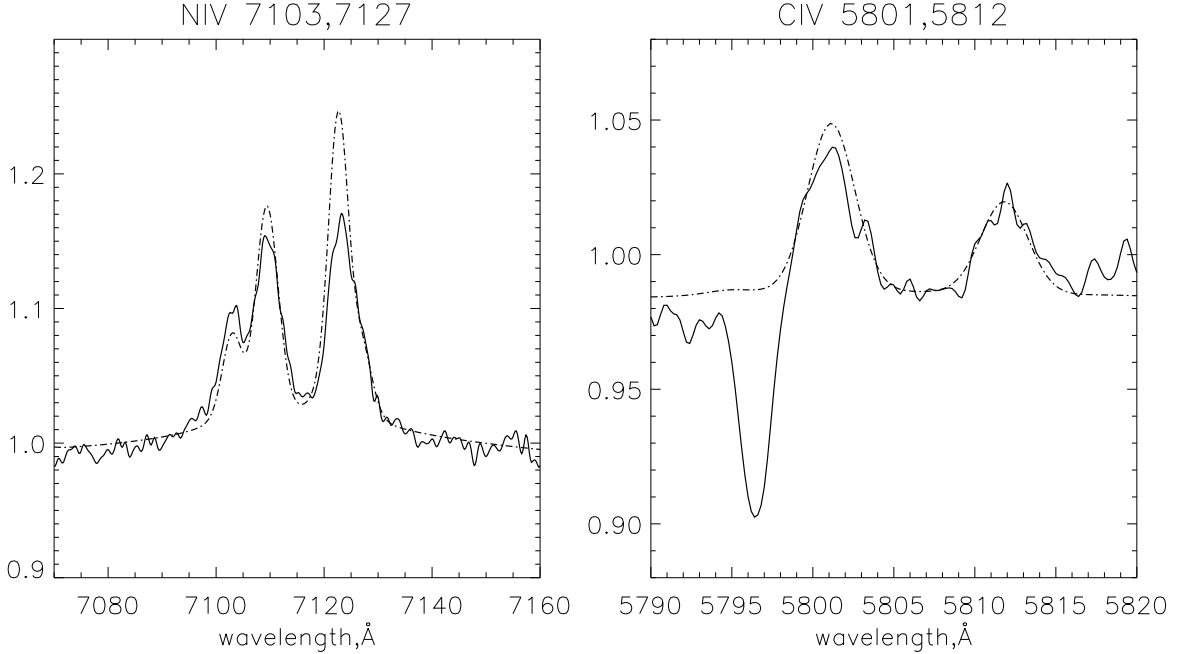,width =1.0\textwidth}
\caption{The profiles of NIV$\lambda\lambda 7103.2-7129.2$ (left) and
      CIV$\lambda \lambda 5801.3, 5812$ (right). Our model ($\beta=2$,
      $\dot{M}=6\times 10^{-7} M_{\odot}/\mbox{year}$) is shown as the
      dot-dashed line. DIB~$\lambda \lambda 5797.03, 5809.10$ appear in
      the observed spectrum.
}
\label{fig:nivmodel}
\end{figure*}

%%%%%%%%%%%%%%%%%%%%%%%%%%%%%%%%%%%%%%%%%%%%%%%%%%%%%%%%%%%%%%%%%%%%%%%%%%%%%%
\begin{table*}%[!p]
%%\setcaptionmargin{0mm} \onelinecaptionstrue
%%\captionstyle{flushleft}
\caption{Derived properties of \objsev. % MWM обозначает $\log(\dot{M}v_\infty R^{0.5})$.
} %Параметры \objsev известные из литературы. }
\label{measurements}
\bigskip
\begin{tabular}{lcccccccc}
\hline
\hline
\label{tab:parmodels}
     & $T_*$  & $R_*$       & $T_{eff}$&$R_{2/3}$&   $ L_*$     & $\dot{M}_{uncl}$        & $v_{\infty}$    & $\beta$ \\% & MWM \\%& Ref \\
     & [kK]   &[$R_{\odot}$]&  [kK]    &  [$R_{\odot}$] &[$10^6 L_{\odot}$]& [$10^{-6}\rm M_{\odot}/year$]& [km/s] &          \\%&    \\%  &     \\
\hline
%    &        &             &          &         &              &                          &                &   \\&&
%    &        &             &          &         &              &                          &                &   \\
Model1        & 45 & 16.5  &  44.5 & 16   &   1                &     7.9         &  3080          &   1   \\ %& &  $29.24\pm0.06$ \\%&   \\% cl =-4.5
{\footnotesize (H$\alpha$)}    &    &       &       &      &                    &                  &                &       \\ %& &                \\%&  \\%&&
%              &    &       &       &      &                    &                  &                &        %& &                \\&&
Model2        & 45 & 16.5  & 44.5  &17    &   1                &     0.95          &  3080          &   2    \\ %&&  $29.75\pm0.05$ \\%&  \\% cl =-4.0
{\footnotesize (NIV, CIV)}   &     &       &       &      &                    &                  &                &        \\ % &              \\%
              &     &       &       &      &                    &                  &                &         \\ %&                 \\%&      \\
$\objsev^*$   & 45.5&  14.6 &       &      &   0.813            &     9.86         &   3080         &   0.9   \\ %& 29.93 \\%&   [1]   \\
%1.000E+06  2.500E-06  4.500E+04  114.430  100.000  4.439E+04  16.893  3.972E-01  1.000E-01  9.660  3250.000  1.000  1.000  2.518E-04   3.791E-05
\hline
\multicolumn{9}{l}{\footnotesize $^*$ -- The data were taken from \citep{Herrero2002}}\\
\hline
\end{tabular}
\end{table*}
%%%%%%%%%%%%%%%%%%%%%%%%%%%%%%%%%%%%%%%%%%%%%%%%%%%%%%%%%%%%%%%%%%%%%%%%%%%%%%
      In order to describe the profiles of the CIV$\lambda \lambda 5801.3, 5812$ and
      NIV~$\lambda\lambda 7103.2-7129.2$ wind lines, we had to construct a model with
      slower acceleration of wind, corresponding to larger values of $\beta$ ($\beta=2$)
      and a lower mass loss rate $\dot{M}_{cl}=(3\pm2)\times 10^{-7}
      M_{\odot}/\mbox{year}$ (Figure~\ref{fig:nivmodel}).

      We now compare our results with earlier studies of  \objsev. \citet{Herrero2002}
      studied its spectrum over a wide range of wavelengths (4000-6700 \AA\AA) using
      the {FASTWIND} code  \citep{fastwind,Puls} which includes the blanketing effect.
      Their results for the H$\alpha$ line are shown on the right panel of
      Figure~\ref{fig:halphamodel}. Table~\ref{tab:parmodels} lists the parameters
      of our models and the parameters obtained by \citet{Herrero2002}. $R_{2/3}$
      is the radius at which the optical depth $\tau$ becomes equal to $2/3 $ and
      $ T_{eff}$ is the effective temperature of the object at $R_{2/3}$ (assuming
      radiative equilibrium). From the table it can be seen that our estimates for
      the  H$\alpha$ line are in good agreement with the earlier measurements
      \citet{Herrero2002}. Note that the nitrogen and carbon emission lines were
      not modelled in \citet{Herrero2002}, so that the differences we have found in
      the parameters derived from the   H$\alpha$ line and the wind lines do not
      contradict earlier results.

%888888888888888888888888888888888 figure 2 88888888888888888888888
%8888888888888888888888888888888888 figure 2 8888888888888888888888

      The rotation of early-type stars with radiatively driven winds leads to interesting
      effects, the most prominent is the tendency to concentrate the outflowing material
      toward regions near the equatorial plane. This results in a deviation from a
      spherically symmetric shape and possibly the formation of an outflowing disk
      \citep{lamers}. Disks of this kind have been discovered in B[e] and Be stars \citep{Be}.
      Asymmetric winds were found in objects that are evolutionary related to O-stars:
      Luminous Blue Variables (LBV) \citep{GrohAGCar,Grohetacar} and Wolf-Rayet stars
      \citep{Harries}. Moreover numerical calculations \citet{GrohqWR} have shown that
      density of the  wind of qWR star HD 45166 varies with latitude. We assume that
      the difference between the model describing  H$\alpha$  and the one describing
      CIV and NIV lines is related to a latitudinal inhomogeneity of the supergiant
      wind due to its rotation ($V \sin{i}=105~\kms$).

%%%%%%%%%%%%%%%%%%%%%%%%%%%%%%%%%%%%%%%%%%%%%%%%%
\section{Conclusions}\label{sec:disc}

      We studied one of the hottest stars in our Galaxy  \objsev by  spectrum obtained
      on the RTT150 telescope. Using non-LTE code { CMFGEN} we estimate its parameters
      (bolometric luminosity, stellar radius, mass loss rate, wind velocity, elementary
      abundances). The atmosphere of this object is rich with nitrogen. We have shown
      that the wind of \objsev is inhomogeneous. Therefore \objsev is yet another star
      with  density of wind depending on latitude.

      The good agreement between the parameters of \objsev found here and those determined
      from spectra over a wide range of wavelengths \citep{Herrero2002} indicates that
      medium-resolution spectra in the red can be used to obtained fairly accurate estimates
      for the parameters of the atmospheres of hot stars when used in combination with
      reliable codes such as { CMFGEN}. Spectra of hot stars at red wavelengths contain lines
      formed in a stellar wind. Thus, when a red-sensitive detector is available,
      medium-resolution spectra can be used to study the strong photospheric lines, as well
      as features of the wind by monitoring its spectral variability.

\bigskip
\bigskip

      O. Maryeva thanks John D. Hillier for his excellent program CMFGEN, used here to analyze
      the data, and S. V. Karpov for help with the calculations. O. M. was supported by the
      ``Kadry'' program (state contract 14.740.11.0800) and the Russian Foundation for Basic
      Research (grants RFFI-11-02-00319-a, 12-07-00739-a). R. Zh. thanks the Russian Foundation for Basic
      Research (grant RFFI-10-02-01145), T\"ubitak, and KFU for partial support in using the RTT-150 telescope.

%%%%%%%%%%%%%%%%%%%%%%%%%%%%%%%%%%%%%%%%%%%%%%%%%%%%%%%%%%%%%%%%%%%%%%%%%%%%%%%%%%%%

 \end{document}